\documentclass{WileyMSP-template}
\usepackage{amsmath}

\begin{document}

\pagestyle{fancy}
\rhead{\includegraphics[width=2.5cm]{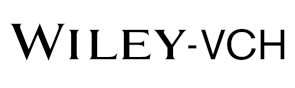}}

\title{Optomechanical cavities based on epitaxial GaP on nominally (001)-oriented Si}

\maketitle


\author{Paula Mouriño$\dagger$}
\author{Laura Mercadé$\dagger$}
\author{Miguel Sinusía Lozano}
\author{Raquel Resta}
\author{Amadeu Griol}
\author{Karim Ben Saddik$\ddagger$}
\author{Enrique Barrigón}
\author{Sergio Fernández-Garrido$\ddagger\ddagger$}
\author{Basilio Javier García}
\author{Alejandro Martínez}
\author{Víctor J. Gómez*}

$\dagger$ Both authors contributed equally.

\begin{affiliations}
P. Mouriño, L. Mercadé, M. Sinusía Lozano, R. Resta, A. Griol, A. Martínez, V. J. Gómez \\
Nanophotonics Technology Center, Universitat Politècnica de València, 46022, Spain.
\\
Email Address: vjgomher@ntc.upv.es

E. Barrigón\\
Departamento Fisica Aplicada I, Universidad de Málaga, 29071, Spain.

K. Ben Saddik, S. Fernández-Garrido, B. J. García\\
Electronics and Semiconductors Group (ElySe), Applied Physics Department, Universidad Autónoma de Madrid, 28049, Spain.

Present address:
$\ddagger$ LAAS-CNRS,7 Avenue du Colonel Roche, 31400 Toulouse, France.

$\ddagger\ddagger$ ISOM and Department of Materials Science, Universidad Politécnica de Madrid, Avda. Complutense 30, 28040 Madrid, Spain.

\end{affiliations}



\begin{abstract} \justifying

Gallium phosphide (GaP) has recently received considerable attention as a suitable material for building photonic integrated circuits due to its remarkable optical and piezoelectric properties. Usually, GaP is grown epitaxially on III-V substrates to keep its crystallinity and later transferred to silicon wafers for further processing. Here, an alternative promising route for the fabrication of optomechanical (OM) cavities on GaP epitaxially grown on nominally (001)-oriented Si is introduced by using a two-step process consisting of a low-temperature etching of GaP followed by selective etching of the underneath silicon. The low-temperature (-30 $^o$C) during the dry-etching of GaP hinders the lateral etching rate, preserving the pattern with a deviation between the design and the pattern in the GaP layer lower than 5 \%, avoiding the complex process of transferring and bonding a GaP wafer to a silicon-on-insulator wafer. To demonstrate the quality and feasibility of the proposed fabrication route, suspended OM cavities are fabricated and experimentally characterized. The cavities show optical quality factors between 10$^3$ and 10$^4$, and localized mechanical resonances at frequencies around 3.1 GHz. Both optical and mechanical resonances are close to those previously reported on crystalline GaP structures. These results suggest a simple and low-cost way to build GaP-based photonic devices directly integrated on industry-standard Si(001) photonic wafers.

\end{abstract}


\section{Introduction} \justifying 
Gallium Phosphide (GaP) displays an interesting combination of a large refractive index (n$_G$$_a$$_P$ = 3.34 at 600 nm \cite{bond_measurement_2004}), wide electronic band-gap (2.26 eV) and large $\chi^2$ and $\chi^3$ nonlinear coefficients that has recently boosted the interest in building nanophotonic and integrated photonic devices using this III-V semiconductor material \cite{wilson_integrated_2020}. The large refractive index of GaP enables tight light confinement in waveguides and nanoresonators. Its wide electronic bandgap produces low two-photon absorption when operating at telecom wavelengths. Furthermore, GaP also displays large $\chi^2$ and $\chi^3$ nonlinear coefficients, which makes it highly suitable for frequency comb generation \cite{wilson_integrated_2020}, harmonic generation \cite{moretti_engineering_2021,mclaughlin_nonlinear_2022}, or all-optical switching \cite{grinblat_ultrafast_2019}. Finally, it is also piezoelectric, so it can eventually be used to build nano-electro-opto-mechanical systems (NEOMS) that, amongst other functionalities, can perform coherent microwave-to-optics transduction mediated by mechanical oscillators \cite{wilson_integrated_2020,schneider_optomechanics_2019,stockill_gallium_2019,honl_microwave--optical_2022,stockill_ultra-low-noise_2022,yama_silicon-lattice-matched_2023,logan_triply-resonant_2023}.

Free-standing cavities have been demonstrated in epitaxial layers of GaP on GaP substrates using an Al$_x$Ga$_1$$_{-}$$_x$P sacrificial layer. This sacrificial layer serves as an etch stop layer and can be selectively wet etched to release free-standing GaP structures \cite{stockill_ultra-low-noise_2022}. However, photonic devices such as waveguides, grating couplers, or ring resonators cannot be fabricated using this approach. A second approach for fabricating GaP photonic integrated circuits (PICs), including optomechanical (OM) cavities, relies on the transfer and bonding of GaP membranes on top of the final substrate. Those membranes can be fabricated via direct wafer bonding of an epitaxial GaP/Al$_x$Ga$_1$$_{-}$$_x$P/GaP wafer structure onto a SiO$_2$-on-Si wafer \cite{wilson_integrated_2020,schneider_optomechanics_2019,logan_triply-resonant_2023,schneider_gallium_2018,honl_highly_2018}. Then, the GaP wafer and Al$_x$Ga$_1$$_{-}$$_x$P sacrificial layer \cite{epple_dry_2002} are sequentially etched, leading to a GaP-on-insulator (GaP-o-I) wafer. After this, the OM cavities are patterned and released by combining anisotropic dry-etching and wet-etching. This process exploits the refractive index contrast between GaP and SiO$_2$ to produce optical confinement. A similar approach has been followed to realize PICs in the GaP-on-diamond platform \cite{chakravarthi_hybrid_2023}. However, the process of transferring and bonding GaP membranes is technologically complex, involves a significant number of steps, and sacrifices a full GaP wafer, which increases the complexity and cost of the process. An alternative and promising approach is growing the GaP layer directly on a silicon (001) wafer instead of transferring the GaP membrane. This approach is, in principle, limited to membrane thicknesses of the order of 40 nm due to the lattice and thermal expansion coefficient mismatch \cite{doscher_situ_2008}. However, advances in the understanding of GaP nucleation on Si(001) \cite{beyer_gap_2012}, together with a high-quality GaP epitaxial growth \cite{beyer_gap_2012,ben_saddik_growth_2021} have overcome the thickness limitation of GaP-on-Si(001) layers. Nonetheless, it is still unclear how the growth of GaP on silicon might affect the OM properties of the cavities. The fabrication of GaP OM cavities directly on silicon presents a highly scalable material platform for photonic device integration.

Here, we demonstrate free-standing GaP OM cavities fabricated on epitaxial GaP-on-Si(001) by low-temperature dry-etching of GaP followed by a highly selective dry-etching of the underneath silicon substrate. The low-temperature (-30 $^o$C) dry-etching of GaP hinders the lateral etching rate of GaP, preserving the pattern with a deviation between the design and the pattern in the GaP layer lower than 5$\%$. During the GaP etching step, the vertical etch rate close to corrugations, walls, and inside holes is dramatically increased due to the inverse aspect ratio dependent etching effect \cite{lai_aspect_2006}, exposing only the silicon surface in those areas. Subsequently, during the selective etching of silicon, the reactive species reach the exposed silicon surfaces, thus starting the etching of silicon underneath the remaining GaP, and finally leading to a suspended GaP OM cavity. The tested cavities show optical quality (Q) factors of the order of 10$^3$ and 10$^4$, and mechanical modes at frequencies around 3.1 GHz, which are values close to those reported in the literature using more complex and expensive fabrication processes \cite{schneider_optomechanics_2019,stockill_gallium_2019,yama_silicon-lattice-matched_2023}. These results prove that our method is a low-cost alternative to integrate GaP-based photonic devices directly on industry-standard Si(001) wafers.

\section{OM crystal cavity design}

\begin{figure}[ht]
    \centering
    \includegraphics[width=0.8\textwidth]{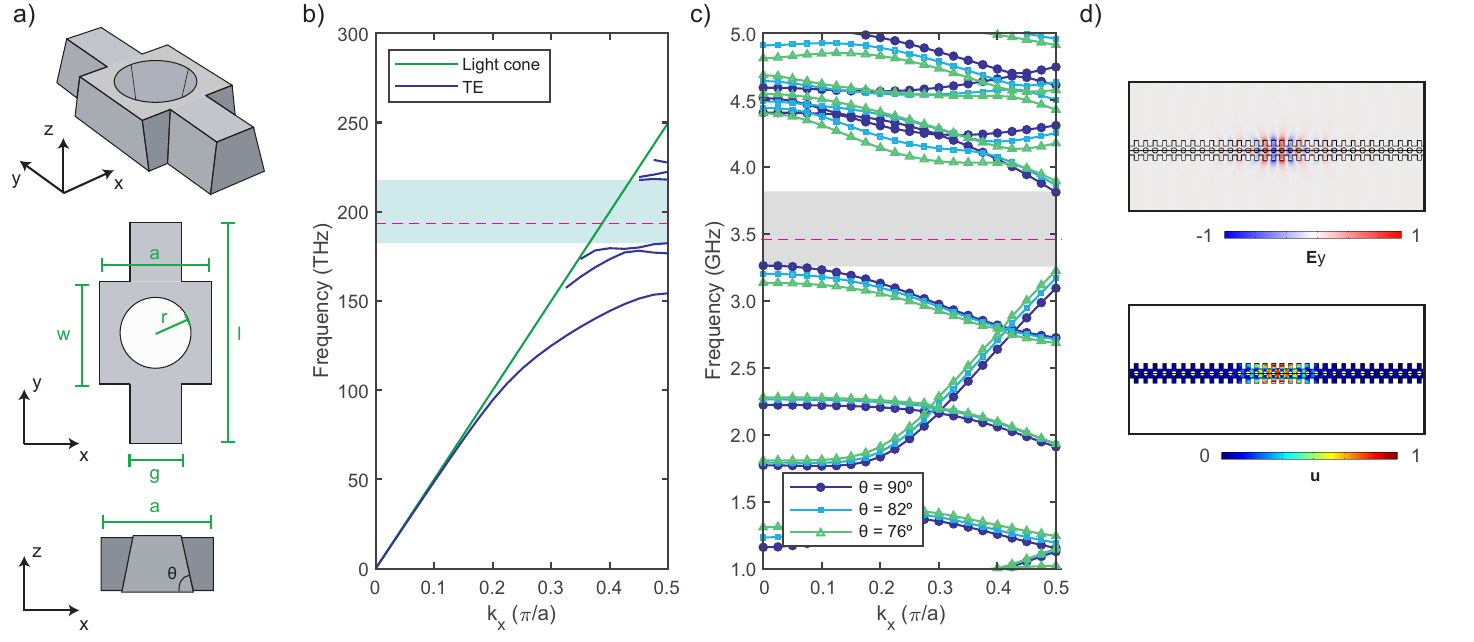}
    \caption{Design of the GaP OM cavity. a) Unit cell geometry depicting the main design parameters considering different etch angles. b) Even-parity (TE) photonic band diagram of the mirror unit cell. c) Mechanical band diagram of the y-symmetric modes showing a mechanical quasi-band gap at the target frequency for different etch angles. In b) and c) the horizontal dashed lines show the predicted resonance frequency, estimated as the frequency of the defect unit cell at the boundary of the Brillouin zone. d) Normalized optical (\textbf{E}$_{y}$) and mechanical (\textbf{u}) field profile giving the highest optomechanical coupling rate.  The unit cell parameters are given in the main text. 
    }
    \label{fig:1}
\end{figure}

We designed the GaP OM cavity starting from a previous cavity demonstrated in silicon \cite{MER21-PRL, MER21-LPOR}, which has the interesting property of displaying a full phononic bandgap at frequencies around 4 GHz. We modified the parameters to account for the GaP refractive index (slightly smaller than that of silicon) as well as a fabrication-induced imperfection that we observed in the fabricated samples: a tilt in the sidewall of the cavity (see Fig. \ref{fig:1}a). By properly choosing the parameters of the unit cell (Fig. \ref{fig:1}a), we can attain a photonic bandgap for TE-like modes (Fig.\ref{fig:1}b) and a phononic bandgap of the y--symmetric bands around 3.5 GHz (Fig.\ref{fig:1}c). Notice that the tilt of the cavity walls can influence the phononic bands. This unit cell can be used to build the mirror of the OM cavity that prevents leakage of both photons and phonons. Once the mirror unit cell has been designed, the cavity can be built by adiabatically changing the unit cell dimensions towards the center of the cavity, according to the process explained in \cite{mercade_microwave_2020}. The parameters of the nominal cavity according to the nomenclature of Fig. \ref{fig:1}a are for the mirror unit cell ($a$,$r$,$g$)$_{m}$=(600,162,240) nm whilst the defect unit cell was built by reducing the design parameter as ($a$,$r$,$g$)$_{d}$=0.6$\times$($a$,$r$,$g$)$_{m}$, the other fixed parameters are a total height of $t$= 240 nm, a total height of $l$= 1250 nm and a width of $w$= 570 nm.  This allows us to get an OM cavity supporting optical and several mechanical resonances, including one at $\approx3.5$ GHz, which exhibits an appreciable OM coupling rate with the optical field (see Fig. \ref{fig:1}c). This value should be large enough to observe the optical transduction of the thermally-populated mechanical modes. The resulting normalized optical (\textbf{E}$_{y}$) and the mechanical (\textbf{u}) field profile with the largest OM coupling rate can be found in Fig. \ref{fig:1}d. The mechanical mode is placed within a phononic quasi-bandgap with a mechanical frequency of 3.48 GHz and an OM coupling rate of g$_{PE+MB}/2\pi$= 1.15 MHz, taking into account both the photoelastic, g$_{PE}/2\pi$= 970 kHz, and the moving boundary effects, g$_{MB}/2\pi$= 184 kHz \cite{PEN14-NP}, both contributions adding up constructively. Notice that photoelastic coefficients of GaP were estimated to be p$_{11}$=-0.0165, p$_{12}$=-0.14 and p$_{44}$=-0.072 \cite{BAL14-OPT}.

\section{Fabrication}
\begin{figure}[ht]
    \centering
    \includegraphics[width=\linewidth]{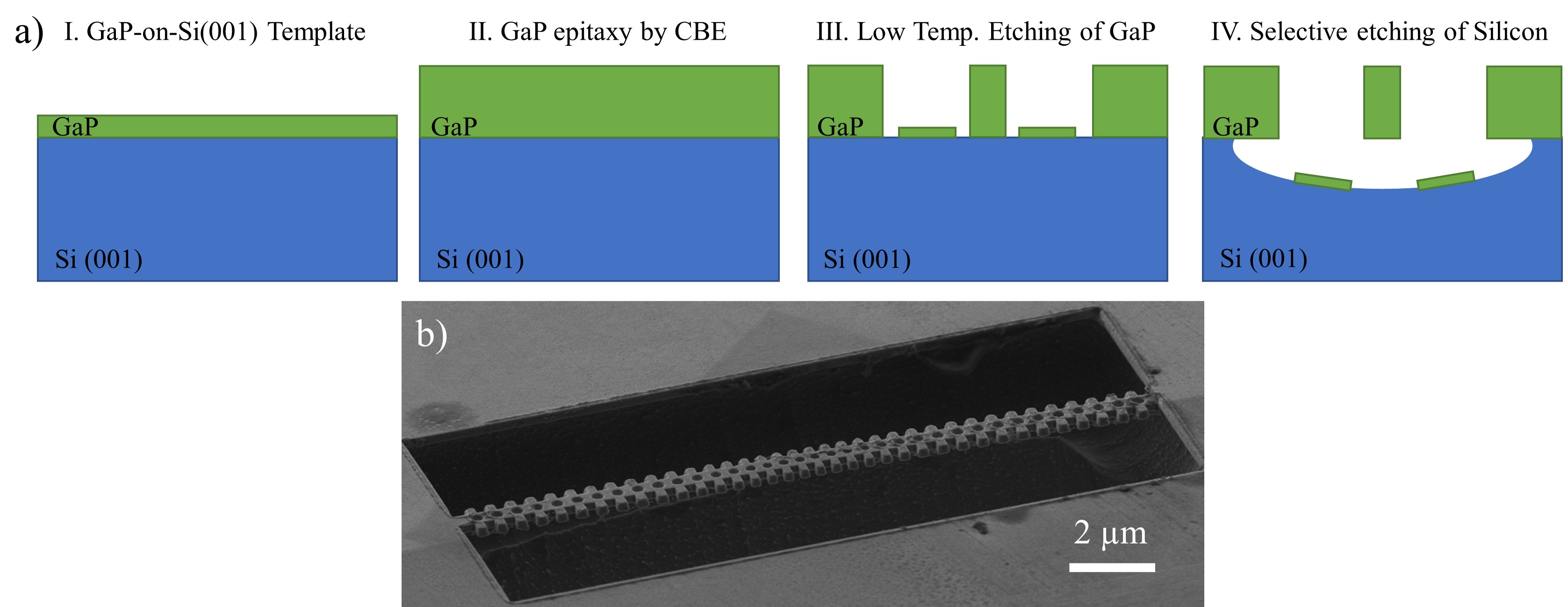}
    \caption{Process flow for the fabrication of GaP samples. a) Schematic of the steps for the fabrication process of OM cavities based on epitaxial GaP-on-Si(001). b) High-resolution SEM micrograph in tilted view at 60$^o$ of a released GaP OM cavity.}
    \label{fig:2}
\end{figure}

The overall fabrication process is schematically illustrated in Fig. \ref{fig:2}a. The process begins with the epitaxial regrowth of a GaP layer on a commercially available GaP-on-Si(001) substrate (steps I and II in Fig. \ref{fig:2}a ) by means of Chemical Beam Epitaxy (CBE) to reach the targeted GaP thickness. The OM cavities are subsequently patterned in the GaP layer utilizing an optimized low-temperature inductively coupled plasma (ICP) reactive ion etching (RIE) process (Fig. \ref{fig:2}a step III). Finally, the cavities are released from the substrate employing a selective dry-etching process (Fig. \ref{fig:2}a step IV). Figure \ref{fig:2}b shows a high-resolution scanning electron microscopy (SEM) image of a GaP OM cavity released from the silicon substrate. The low-temperature etching of GaP and the selective etching of silicon steps are described in detail in the following subsections.

\subsection{Low-temperature etching of GaP}

The patterning of the OM cavities in the top GaP layer is crucial. In general, photonic devices such as grating couplers, waveguides, and ring resonators require well-defined (as vertical as possible) and smooth sidewalls to minimize scattering losses. Particularly, the fabrication of OM cavities needs a high-resolution process capable of transferring a plethora of features with sizes ranging from tenths to hundreds of nanometers and high dimensional accuracy, i.e., the deviation between the design and the pattern in the GaP layer should be minimum \cite{schneider_strong_2016,seidler_optimized_2017}. Schneider and co-workers \cite{schneider_optomechanics_2019} demonstrated that the optical quality factor drops by 70$\%$ if the hole radii are 10$\%$ larger than those of the design.\\

For the anisotropic dry-etching of GaP, either a chlorine-based chemistry, such as Cl$_2$ or BCl$_3$ \cite{lee_plasma_1997,shul_high-density_1997,smolinsky_plasma_1981}, mixtures of H$_2$ and CH$_4$ \cite{pearton_comparison_1996,pearton_high_1996}, or a combination of both \cite{shul_temperature_1996} have been typically employed. For a more physical dry-etching, Ar and N$_2$ gases are usually included in the mixture \cite{lee_plasma_1997,shul_high-density_1997,yang_experimental_2007}. Analyzing the interplay of the different process parameters is highly complicated, but we can extract some general trends. A plasma with a high density of chlorine atoms and ions achieves etch rates exceeding 1500 nm/min \cite{shul_high-density_1997}. High-density Cl$_2$-based plasmas have been demonstrated to achieve GaP etch rates over 1500 nm/min due to the high density of Cl neutrals and ions generated as compared to BCl$_3$-based plasmas. The sample electrode temperature has a significant influence on the sidewall profile, with lower temperatures leading to reduced undercut and roughness, most likely due to the reduction of the volatility of GaCl$_x$ and GaF$_x$ by-products that are adsorbed at the sidewalls producing passivation effects \cite{honl_highly_2018,shul_temperature_1996}, CH$_4$ is expected to play an important role \cite{shul_high-density_1997}. Chemically inert gases, such as Ar, contribute to the etch process in a purely physical manner. Keeping the previous points in mind, we developed the ICP-RIE dry-etch process employed in this fabrication process comprising a mixture of 25 sccm Cl$_2$, 8 sccm CHF$_3$, and 70 sccm Ar at a pressure of 5 mTorr, ICP power of 400 W, RF power of 70 W and 1 minute. The selectivity of this etching process over the PMMA resist is 1:1.\\

\begin{figure}[ht]
    \centering
    \includegraphics[width=0.5\textwidth]{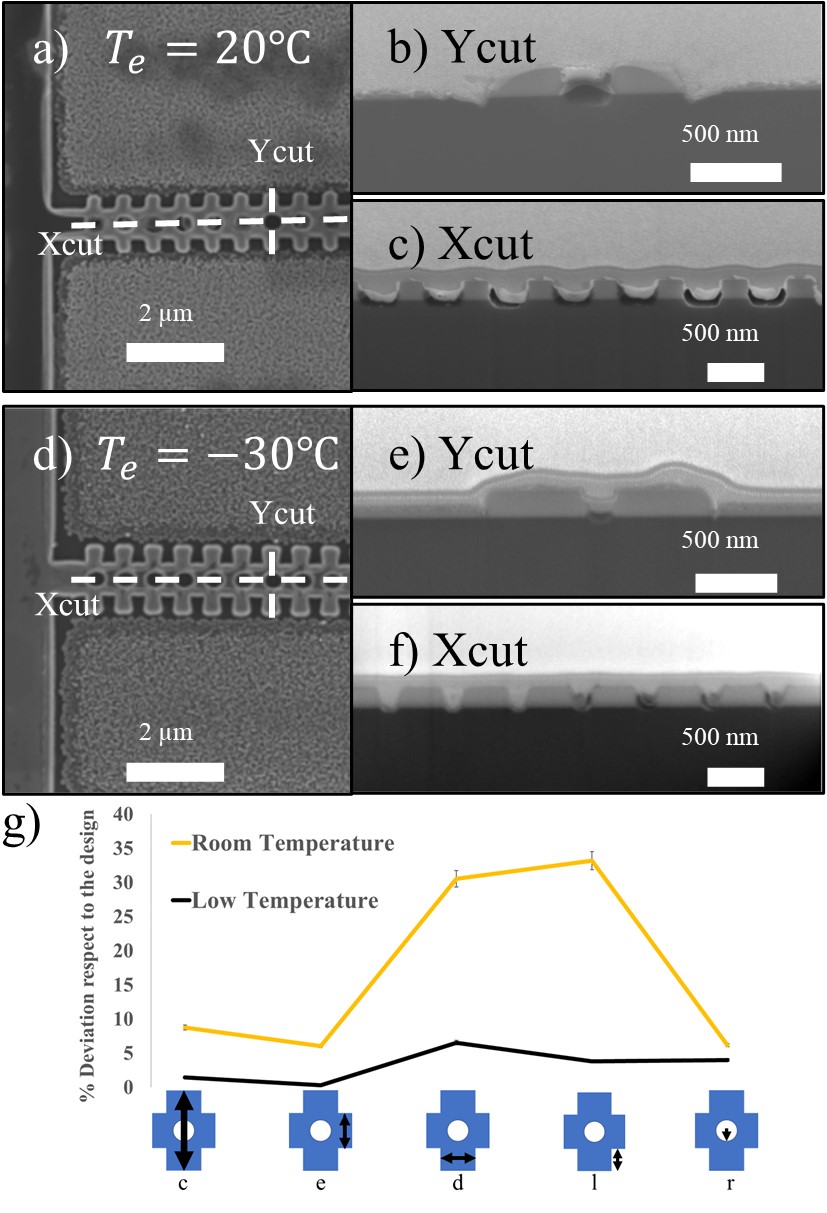}
    \caption{SEM micrographs in top-view, and SEM micrographs of the cross-sections prepared by FIB milling of the cavities etched at a)-c) 20 $^o$C (RT) and d)-f) -30 $^o$C (LT). g) Percentage of deviation of the pattern in GaP with respect to the design after the etching of GaP layer at 20 $^o$C (RT), and -30 $^o$C (LT).}
    \label{fig:3}
\end{figure}

Figure \ref{fig:3} shows the OM cavities experimentally transferred to the GaP layer following the dry-etching process previously described. The dry-etching of GaP was studied for two different etching temperatures of 20 $^o$C (Fig. \ref{fig:3}a to c) and -30 $^o$C (Fig. \ref{fig:3}d to f). Herein, we will refer to them as room-temperature (RT) and low-temperature (LT) recipes, respectively. The energy dispersive x-ray (EDS) maps for the RT and LT recipes can be found in the supporting information (Fig. S1). The vertical etch rates of GaP are 174 $\pm$ 9 nm$\cdot$min$^-$$^1$ and 153 $\pm$ 8 nm$\cdot$min$^-$$^1$ for the RT and LT recipes, respectively. Meanwhile, the estimated etch rates are 50 $\pm$ 20 nm$\cdot$min$^-$$^1$ and 10 $\pm$ 5 nm$\cdot$min$^-$$^1$ in the lateral direction for the RT and LT recipes, respectively. Figure 3a and Figure 3d show the SEM micrographs in the top view for the RT and LT recipes. The dashed lines named Xcut and Ycut specify where the FIB cuts took place. The SEM micrographs of the cross sections prepared by FIB milling with Ga ions are shown in Figure 3, the RT (Fig. \ref{fig:3}b and c) and LT (Fig. \ref{fig:3}e and f) recipes. The sidewalls were analyzed using the cross sections prepared by FIB inside the holes and in the lateral walls of the OM cavities. The sidewall angles were 82 $\pm$ 1 $^o$ (RT recipe) and 76 $\pm$ 1 $^o$ for the LT recipe. Moreover, the Ycut (Fig. \ref{fig:3}b and e) and Xcut (Fig. \ref{fig:3}c and f) micrographs reveal that there is a complete removal of the III-V material inside the holes or close to the GaP sidewalls, and the etching of the silicon substrate is started. The etching rate close to the sidewalls or inside the holes is approximately two times faster than in the flat areas of GaP. The enhanced etch rate in small features (holes and corrugations in our case) is a well-known effect called inverse aspect ratio dependent etching (ARDE)\cite{lai_aspect_2006}. This effect appears when etching high-aspect ratio features using etch inhibitor layers. In high-aspect ratio features, the deposition of etch inhibitor layers is lower than in low-aspect-ratio features due to the shadowing of the precursor species of the etch inhibitor layer. This leads to a lower etching inhibition effect in features with a lower amount of etch inhibitor. This results in a higher etch rate in high-aspect-ratio features. The formation of sloped sidewalls can be attributed to the reflection on ions at the feature sidewall slope. This effect is related to the incoming ion angular distribution as well as the angular distribution of the reflected ions \cite{abdollahi-alibeik_analytical_1999}.\\

Schneider and co-workers \cite{schneider_optomechanics_2019} demonstrated the importance of pattern preservation for the optical performance of the OM cavities, where deviations from the design can be detrimental to the quality factor. To select the appropriate etching conditions for our process, we quantify the deviation (\%) between the pattern transferred to the GaP and the design of the OM cavity as shown in section 2. Figure \ref{fig:3}g represents this deviation as a percentage for the RT and LT recipes for five different parameters of the OM cavities. In general, the RT recipe gives deviations larger than 10\%, being $>$30\% for the $d$ and $l$ parameters. Meanwhile, the deviations are lower than 5\% for the LT recipe. In particular, the dimensions $c$ and $e$ show a deviation lower than 1\% when etching with the LT recipe. For example, for the $c$ parameter, the sizes are 114 $\pm$ 2 nm for the RT recipe and 127 $\pm$ 2 nm for the LT recipe. This means a deviation of 9\% and 1\%, respectively, with respect to the design. It can be appreciated that the etching temperature plays an important role not only in controlling the vertical etch rate but also in influencing the lateral etch rate. As mentioned before, the formation of nonvolatile protective layers reduces the etch rate of the surface covered by the ad-layer. In general, there are two main factors governing the formation of etch inhibitor layers: i.- the species present in the gas mixture \cite{schneider_gallium_2018,shul_temperature_1996}; and ii.- the etching temperature \cite{shul_temperature_1996}. In this process, the etching action is attributed primarily to chlorine-containing species (Cl$_2$ in this case) together with the Ar ions, whereas the fluorine-containing (CHF$_3$ in this case) species are required for the formation of relatively nonvolatile GaF$_x$ layer that serves as an etch inhibitor. Etch inhibitor layers are key to anisotropy and are often used to suppress the etching in the lateral direction and to maintain high etch rates in the vertical direction. On the other hand, our experiments revealed a drastic reduction in lateral etch rate when working with the LT recipe. This can be attributed to the reduction in the volatility of the GaF$_x$ layer with an increase in the etching temperature \cite{honl_highly_2018,shul_temperature_1996}.

\subsection{Selective etching of silicon}
\begin{figure}[ht]
    \centering
    \includegraphics[width=0.8\textwidth]{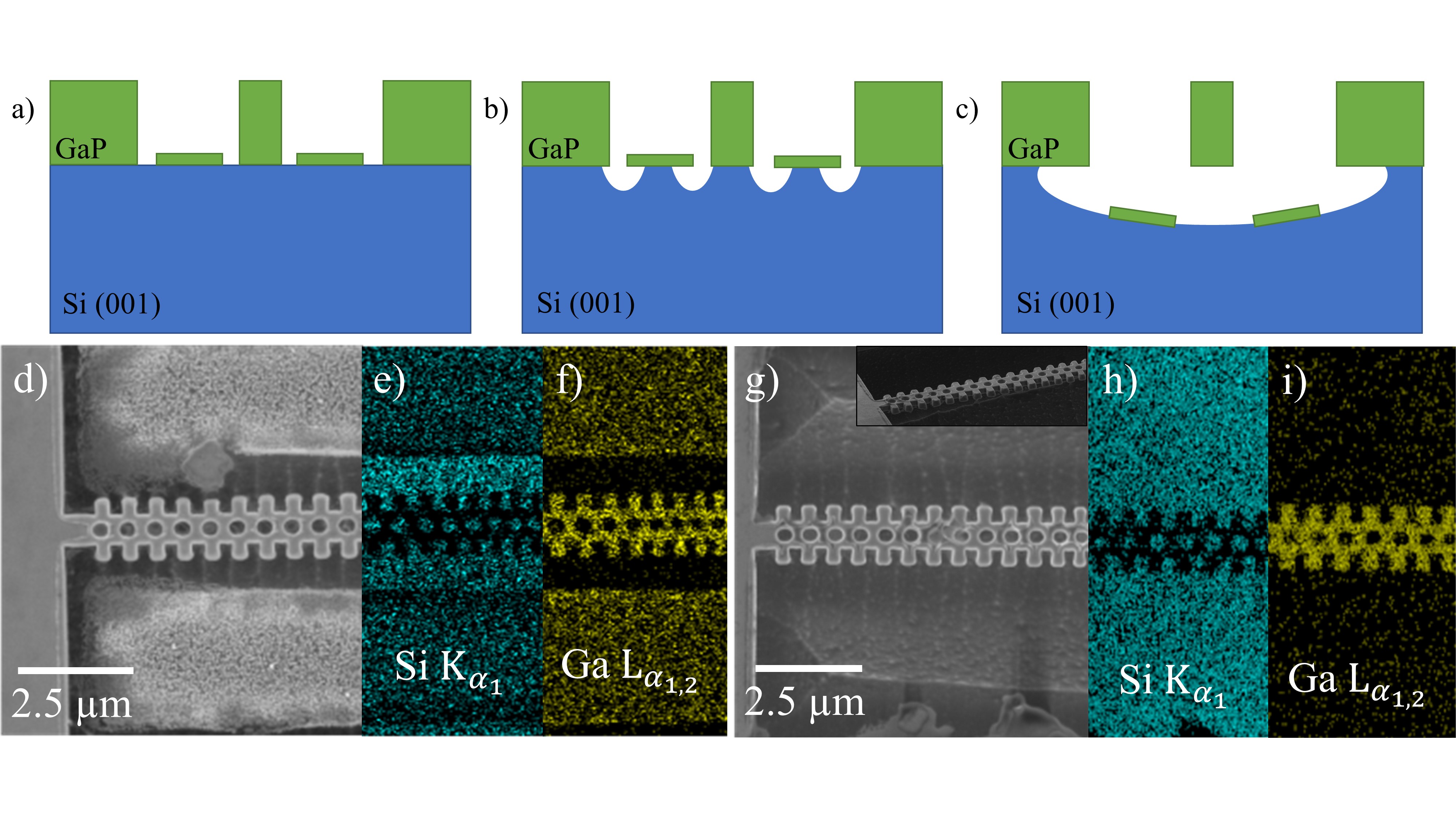}
    \caption{Schematic representation of the selective etching of silicon a) after the GaP etching, b) at the beginning of the selective etching of silicon, and c) after releasing the OM cavity. High-resolution SEM micrographs and EDS maps (top view) of a GaP OM cavity after 8 minutes d)-f) and 16 minutes g)-i), respectively. The inset of panel g) shows the released OM cavity.}
    \label{fig:4}
\end{figure}

After the patterning of the GaP layer, the OM cavities are released to form freestanding structures in a final step by a selective dry etching process of the underlying Si substrate. Selective etching of Si in the presence of GaP was provided by the second and final step, which consisted of another ICP-RIE process (Fig. \ref{fig:4}). The only previously known selective etching process employs XeF$_2$ vapor. This was successfully applied for releasing GaP OM cavities by Yama et al. \cite{yama_silicon-lattice-matched_2023}. As XeF$_2$ vapor is not readily available in all laboratories, we investigated an alternative route for the selective etching of silicon based-on a more widespread gas. In this case, we developed an ICP-RIE process to etch the underneath silicon at room temperature utilizing 50 sccm SF$_6$, at a pressure of 20 mTorr, ICP power of 1000 W, RF power off, and 16 minutes. The silicon substrate was etched at a rate of 200 nm/min, and the selectivity over GaP was estimated to be approximately 200:1. It is worth mentioning that the root mean square roughness of the surface of GaP increases from 0.7 nm to 0.9 nm after the selective etching of silicon.

As described in the previous sub-section, during the GaP etching step, the vertical etch rate close to the corrugations, walls, and inside the holes is more than 1.5 times larger than in the flat opening areas. Therefore, the silicon surface is partially exposed. The rate of chemical species (SF$_6$) reaching the silicon surface is limited by the area of the Si surface exposed (see Fig. \ref{fig:4}b). This results in a slower silicon etch rate at the beginning of the process. This step is characterized by top-view SEM (Fig. \ref{fig:4}d) and EDS maps (Fig. \ref{fig:4}e and f) of the GaP OM cavity after 8 minutes of selective Si etching step. In the EDS maps of the GaP OM cavity, it can be appreciated the formation of a trench on both sides of the OM cavity, confirmed by the higher intensity of the Si K$_\alpha$$_1$ line (Fig. \ref{fig:4}e)) and the lower intensity of the Ga L$_\alpha$$_1$$_,$$_2$ line (Fig. \ref{fig:4}f) close to the OM cavity. This trench extends 630 $\pm$ 28 nm from the end of the corrugations of the OM cavity. 
As the process evolves, the silicon surface is exposed in both the vertical and lateral directions. This way, the amount of reactive species reaching the silicon surface increases with time, resulting in an increase in the etch rate as the process proceeds. This step is characterized using top-view SEM (Fig. \ref{fig:4}g) and energy EDS maps (Fig. \ref{fig:4}h and f) of the GaP OM cavity after 16 minutes of a selective Si etching step. After 16 minutes the trenches have completely disappeared, leading to the full release of the GaP OM cavity. The EDS maps confirm the release of the OM cavity. The Si K$_\alpha$$_1$ line (Fig. \ref{fig:4}h) is found only at the bottom of the structure, while the Ga L$_\alpha$$_1$$_,$$_2$ line (Fig. \ref{fig:4}i) can be found only in the GaP OM cavity, confirming that the cavity is fully suspended. The full release of the OM cavities is also confirmed by SEM in 60$^o$ tilted view, as shown in the inset of Figure \ref{fig:4}g.

\section{Device characterization}
\begin{figure}[h]
    \centering
    \includegraphics[width=0.8\textwidth]{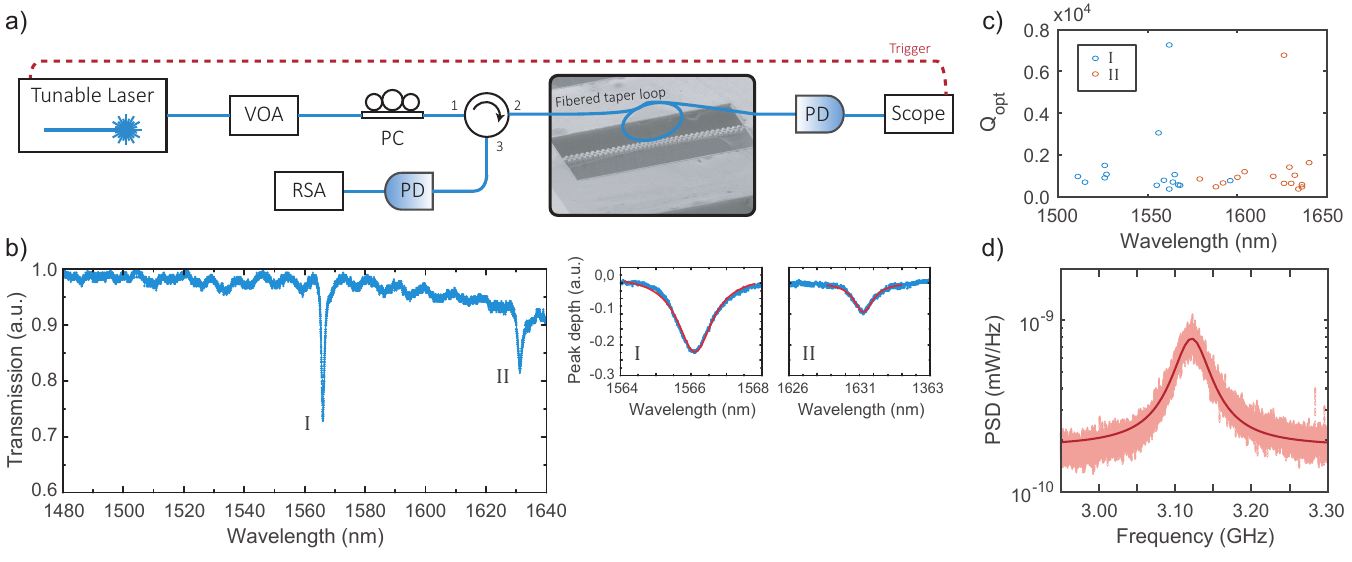}
    \caption{Experimental measurement of the fabricated cavities. a) Optomechanical characterization setup for the optical and mechanical spectra acquisition. The SEM micrograph corresponds to sample S1. b) Optical transmission spectra showing two optical resonances. c) Optical quality factors as a function of the center wavelength for the two optical resonances of a set of tested cavities. d) Power spectral density of the transduced mechanical mode acquired in the reflection channel for one of the tested OM cavities.}
    \label{fig:5}
\end{figure}

Following the route studied in sections 3.1 and 3.2, we prepared two samples. Sample S1: 1 minute LT-GaP etching followed by 16 minutes of selective etching of silicon (Fig. S2a and b) and sample S2: 2 minutes LT-GaP etching followed by 8 minutes of selective etching of silicon (Fig. S2c and d), respectively. It can be appreciated a dramatic increase in the surface roughness for sample S2.
The OM characterization has been performed with the experimental setup shown in Fig. \ref{fig:5}a. Here, a tunable laser is fed into an OM cavity using a fibered taper loop \cite{mercade_microwave_2020} to couple light into and out of the cavity. The optical driving power is controlled using a variable optical attenuator (VOA), and the polarization is properly adjusted through a polarization controller (PC). The transmitted signal is then sent to a photodetector (PD) and recorded with a digital oscilloscope. The reflection channel, enabled by an optical circulator, is photodetected and analyzed with a signal analyzer (SA).

The optical transmission spectrum of one of the tested cavities can be seen in Fig. \ref{fig:5}b. Here, two optical resonances can be found, one around the design wavelength of 1566 nm with an optical quality factor (Q$_{opt}$) of 1.32$\times$10$^3$ at this specific cavity as in the nominal design and another one at 1624 nm and Q$_{opt}$=0.96$\times$10$^3$. The insets in Fig. \ref{fig:5}b show the peak depth of both resonances, where the baseline has been removed to be able to perform a Lorentzian fit to retrieve the wavelength and quality factor parameters. It has to be noted, however, that fabrication imperfections could lead to a shift in the optical resonances, as can be seen in Fig. \ref{fig:5}c, which represents the measured optical Q factor of both resonances for a set of 15 fabricated cavities with the same nominal values. By exciting the cavity with the first optical mode, the mechanical mode is transduced on the optical signal and can be recorded after photodetecting the reflection channel. The resulting power spectral density (PSD) can be found in Fig. \ref{fig:5}d at a target frequency of 3.12 GHz with a mechanical linewidth of $\Gamma_{m}/2\pi$= 49.4 MHz. This results in a mechanical quality factor lower than in previously tested GaP cavities \cite{wilson_integrated_2020,schneider_optomechanics_2019,stockill_gallium_2019,honl_microwave--optical_2022}. The larger mechanical losses in our cavities can be due either to surface effects due to the etching process or to the properties of the bulk GaP material itself. In further work, we plan to build and test more cavities, including designs at other frequencies, so that we can get more information about the origin of the mechanical losses and identify ways to reduce them.

\section{Conclusion}
In summary, we have demonstrated a new fabrication procedure to build released GaP OM cavities on epitaxially grown GaP on nominally (001)-oriented Si substrates by low-temperature and selective etching. In contrast to previous works, our method avoids the process of transferring and bonding GaP membranes from the growth substrate to the photonic wafer. This reduces the number of processing steps and avoids the full sacrifice of III-V wafers. The low-temperature (-30 $^o$C) dry-etching of GaP hinders the lateral etching rate of GaP, preserving the pattern with a deviation between the design and the pattern transferred to the GaP layer lower than 5$\%$. We have found that during the GaP etching step, the inverse-ARDE effect \cite{lai_aspect_2006} is responsible for a dramatic increase in the vertical etch rate close to corrugations, walls, and inside holes, exposing only the silicon surface in those areas. For the selective etching of silicon, we have developed a new process that, up to the best of our knowledge, has not been reported previously. During the selective etching of silicon, the reactive species reach the exposed silicon surfaces at a rate proportional to the exposed surface. This starts the etching of silicon underneath the remaining GaP and finally leads to a suspended GaP cavity. To test the validity of this approach, we have successfully fabricated GaP nanobeam OM cavities, which show optical Q-factors between 10$^3$ and 10$^4$, and mechanical modes at frequencies around 3.1 GHz. Our method could be considered a low-cost alternative to integrate GaP-based photonic devices directly on industry-standard Si(001) wafers.


\section{Experimental section}
\threesubsection{Epitaxial growth of GaP}\\
The GaP layer was primarily grown by chemical beam epitaxy (CBE), using a Riber CBE32 system manufactured by Riber, France, on 3 x 2 cm$^2$ GaP-on-Si(001) substrate diced from a 12 inches wafer purchased from NAsP$_{III/V}$ GmbH, Germany. The GaP-on-Si(001) wafer is nominally exact (001) orientated with a slight miscut of $\approx$0.3$^o$ toward one of the four <110> directions. The GaP buffer layer, purchased from NAsP$_{III/V}$ GmbH, is 25 nm thick and it was prepared by metal-organic chemical vapor deposition (MOCVD). This buffer layer is free of dislocations, stacking faults, and twins, and possesses a smooth surface (root mean square roughness of 0.3 nm) without anti-phase domains. For the growth of the GaP layer by CBE, the as-received substrate was In-bonded onto a Mo holder and outgassed inside the growth chamber at 610 $^o$C to desorb the native oxide. The GaP layer was grown using CBE using tertiarybutylphosphine (TBP) at fluxes that correspond, in equivalent growth rate units of monolayers per second, to 0.5 and 0.22 ML/s, respectively. TBP and TEGa gas precursors were injected into the growth chamber using high-temperature (820 $^o$C) and low-temperature (120 $^o$C) gas injectors, respectively. The thickness of the GaP layer prepared by CBE is 245 nm. Last, for the subsequent fabrication of the OM cavities, the resulting GaP on Si sample was diced into 0.5 x 0.5 cm$^2$ pieces. More details can be found in the work of Saddik \textit{et al}. \cite{ben_saddik_growth_2021}.

\threesubsection{Electron Beam Lithography (EBL)}\\
All cavity designs were patterned on polymethyl methacrylate (PMMA) resist by EBL in an e-beam RAITH 150 system manufactured by RAITH, Germany. The voltage of the column was set at 10 kV and the aperture at 30 $\mu$m.

\threesubsection{Low-temperature Dry-etching of GaP}\\
After patterning and developing the resist, the patterns were transferred to the GaP layer using a low-temperature ICP-RIE process carried out in a Corial 210IL – 200 mm ICP-RIE system manufactured by Corial, France. The pattern created on the resist layer was transferred further into the GaP layer with a mixture of 25 sccm Cl$_2$, 8 sccm CHF$_3$, and 70 sccm Ar at a pressure of 5 mTorr, ICP power of 400 W, RF power of 70 W and 1 minute. The etching process was characterized for two different temperatures of the substrates: 20 $^o$C and -30 $^o$C. The samples were placed at the center of a 200 mm Si carrier wafer with carbon tape for better adhesion without using any thermal conduction promoter such as oil, wax, or grease to improve the thermal contact between the unpolished backside of the chips and the carrier wafer. 

\threesubsection{Selective etching of silicon}\\
After the etching of the GaP layer, the silicon underneath was selectively etched in an STS Multiplex ICP System manufactured by STS, United Kingdom. The holes and trenches etched into the GaP layer allowed the reactive species to reach the silicon surface. The underneath silicon was etched at room temperature with 50 sccm SF$_6$ at a pressure of 20 mTorr, ICP power of 1000 W, RF power off, and 16 minutes. 

\threesubsection{Scanning Electron Microscopy (SEM) characterization}\\
The SEM micrographs were collected using a field-emission scanning electron microscope Gemini SEM500, manufactured by Zeiss, Germany, operated at 3.5 kV. For the measurement of dimensions and profiles of the cavities, the ImageJ\textsuperscript{\textregistered} software was employed \cite{schneider_nih_2012}. Tilted micrographs were acquired at 60$^o$.

\threesubsection{Focused Ion Beam Milling (FIB)}\\
The FIB milling experiments were performed in a Helios Nanolab 650 Dual Beam, manufactured by FEI Company, United States, equipped with a Schottky field emission source for SEM (FESEM) and a Tomahawk focused ion beam (FIB). Different areas of the sample were analyzed by preparing cross-sections using focused ion beam scanning electron microscopy (FIB-SEM) milling (at 30 kV and 0.23 nA-80pA). The surface was previously protected by a layer of platinum deposited by GIS of Pt at 30 kV and 0.23 nA from FIB and 2 kV and 0.8 nA from SEM. The cross-section SEM micrographs were taken at 5 kV.

\medskip
\textbf{Supporting Information} \par
Supporting Information is available from the Wiley Online Library or from the author.

\medskip
\textbf{Acknowledgements} \par 
\begin{sloppypar}
V.J.G., P.M.M, and R.R.L acknowledge financial support from the Generalitat Valenciana (Project: INVEST/2022/170). M.S.L. acknowledges financial support from the Generalitat Valenciana (Project: CIAPOS/2021/293). V.J.G. acknowledges financial support from the Generalitat Valenciana (CDEIGENT/2020/009) and to the AGENCIA ESTATAL DE INVESTIGACIÓN of Ministerio de Ciencia e Innovacion (CNS2023-145093). S.F.G. acknowledges the financial support received through the program Ramón y Cajal (co-financed by the European Social Fund)
under Grant No. RYC-2016-19509. All authors acknowledge financial support from AGENCIA ESTATAL DE INVESTIGACIÓN of Ministerio de Ciencia e Innovacion (PID2020-118855RB-I00, PID2020-114280RB-I00, and PID2021-124618NB-C21) and to the European Regional Development Fund (ERDF) (IDIFEDER/2020/041, IDIFEDER/2021/061). This study formed part of the Quantum Communications program and was supported by Ministerio de Ciencia e Innovacion (MCIN) with funding from European Union NextGenerationEU (PRTR-C17.I1) and by Generalitat Valenciana (COMCUANTICA/003 and COMCUANTICA/004).
\end{sloppypar}

\medskip
\textbf{Conflict of Interest} \par
The authors declare no conflict of interest.

\medskip

%

\bibliographystyle{MSP}
\bibliography{main}

\begin{thebibliography}{10}
\providecommand{\url}[1]{\texttt{#1}}
\providecommand{\urlprefix}{URL }

\bibitem{bond_measurement_2004}
W.~L. Bond,
\newblock \emph{Journal of Applied Physics} \textbf{2004}, \emph{36}, 5 1674.

\bibitem{wilson_integrated_2020}
D.~J. Wilson, K.~Schneider, S.~Hönl, M.~Anderson, Y.~Baumgartner, L.~Czornomaz, T.~J. Kippenberg, P.~Seidler,
\newblock \emph{Nature Photonics} \textbf{2020}, \emph{14}, 1 57.

\bibitem{moretti_engineering_2021}
G.~Q. Moretti, E.~Cortés, S.~A. Maier, A.~V. Bragas, G.~Grinblat,
\newblock \emph{Nanophotonics} \textbf{2021}, \emph{10}, 17 4261.

\bibitem{mclaughlin_nonlinear_2022}
B.~McLaughlin, D.~P. Lake, M.~Mitchell, P.~E. Barclay,
\newblock \emph{Journal of the Optical Society of America B} \textbf{2022}, \emph{39}, 7 1853.

\bibitem{grinblat_ultrafast_2019}
G.~Grinblat, M.~P. Nielsen, P.~Dichtl, Y.~Li, R.~F. Oulton, S.~A. Maier,
\newblock \emph{Science Advances} \textbf{2019}, \emph{5}, 6 34.

\bibitem{schneider_optomechanics_2019}
K.~Schneider, Y.~Baumgartner, S.~Hönl, P.~Welter, H.~Hahn, D.~J. Wilson, L.~Czornomaz, P.~Seidler,
\newblock \emph{Optica} \textbf{2019}, \emph{6}, 5 577.

\bibitem{stockill_gallium_2019}
R.~Stockill, M.~Forsch, G.~Beaudoin, K.~Pantzas, I.~Sagnes, R.~Braive, S.~Gröblacher,
\newblock \emph{Physical Review Letters} \textbf{2019}, \emph{123}, 16 163602.

\bibitem{honl_microwave--optical_2022}
S.~Hönl, Y.~Popoff, D.~Caimi, A.~Beccari, T.~J. Kippenberg, P.~Seidler,
\newblock \emph{Nature Communications} \textbf{2022}, \emph{13}, 1 2065.

\bibitem{stockill_ultra-low-noise_2022}
R.~Stockill, M.~Forsch, F.~Hijazi, G.~Beaudoin, K.~Pantzas, I.~Sagnes, R.~Braive, S.~Gröblacher,
\newblock \emph{Nature Communications} \textbf{2022}, \emph{13}, 1 6583.

\bibitem{yama_silicon-lattice-matched_2023}
N.~S. Yama, I.-T. Chen, S.~Chakravarthi, B.~Li, C.~Pederson, B.~E. Matthews, S.~R. Spurgeon, D.~E. Perea, M.~G. Wirth, P.~V. Sushko, M.~Li, K.-M.~C. Fu,
\newblock \emph{Advanced Materials} \textbf{2023}, \emph{36}, 5 2305434.

\bibitem{logan_triply-resonant_2023}
A.~D. Logan, S.~Shree, S.~Chakravarthi, N.~Yama, C.~Pederson, K.~Hestroffer, F.~Hatami, K.-M.~C. Fu,
\newblock \emph{Optics Express} \textbf{2023}, \emph{31}, 2 1516.

\bibitem{schneider_gallium_2018}
K.~Schneider, P.~Welter, Y.~Baumgartner, H.~Hahn, L.~Czornomaz, P.~Seidler,
\newblock \emph{Journal of Lightwave Technology} \textbf{2018}, \emph{36}, 14 2994.

\bibitem{honl_highly_2018}
S.~Hönl, H.~Hahn, Y.~Baumgartner, L.~Czornomaz, P.~Seidler,
\newblock \emph{Journal of Physics D: Applied Physics} \textbf{2018}, \emph{51}, 18 185203.

\bibitem{epple_dry_2002}
J.~H. Epple, C.~Sanchez, T.~Chung, K.~Y. Cheng, K.~C. Hsieh,
\newblock \emph{Journal of Vacuum Science \& Technology B: Microelectronics and Nanometer Structures Processing, Measurement, and Phenomena} \textbf{2002}, \emph{20}, 6 2252.

\bibitem{chakravarthi_hybrid_2023}
S.~Chakravarthi, N.~S. Yama, A.~Abulnaga, D.~Huang, C.~Pederson, K.~Hestroffer, F.~Hatami, N.~P. de~Leon, K.-M.~C. Fu,
\newblock \emph{Nano Letters} \textbf{2023}, \emph{23}, 9 3708.

\bibitem{doscher_situ_2008}
H.~Döscher, T.~Hannappel, B.~Kunert, A.~Beyer, K.~Volz, W.~Stolz,
\newblock \emph{Applied Physics Letters} \textbf{2008}, \emph{93}, 17 172110.

\bibitem{beyer_gap_2012}
A.~Beyer, J.~Ohlmann, S.~Liebich, H.~Heim, G.~Witte, W.~Stolz, K.~Volz,
\newblock \emph{Journal of Applied Physics} \textbf{2012}, \emph{111}, 8 083534.

\bibitem{ben_saddik_growth_2021}
K.~Ben~Saddik, B.~J. García, S.~Fernández-Garrido,
\newblock \emph{APL Materials} \textbf{2021}, \emph{9}, 12 121101.

\bibitem{lai_aspect_2006}
S.~L. Lai, D.~Johnson, R.~Westerman,
\newblock \emph{Journal of Vacuum Science \& Technology A} \textbf{2006}, \emph{24}, 4 1283.

\bibitem{MER21-PRL}
L.~Mercad\'e, K.~Pelka, R.~Burgwal, A.~Xuereb, A.~Mart\'{\i}nez, E.~Verhagen,
\newblock \emph{Phys. Rev. Lett.} \textbf{2021}, \emph{127} 073601.

\bibitem{MER21-LPOR}
L.~Mercadé, M.~Morant, A.~Griol, R.~Llorente, A.~Martínez,
\newblock \emph{Laser \& Photonics Reviews} \textbf{2021}, \emph{15}, 11 2100175.

\bibitem{mercade_microwave_2020}
L.~Mercadé, L.~L. Martín, A.~Griol, D.~Navarro-Urrios, A.~Martínez,
\newblock \emph{Nanophotonics} \textbf{2020}, \emph{9}, 11 3535.

\bibitem{PEN14-NP}
Y.~Pennec, V.~Laude, N.~Papanikolaou, B.~Djafari-Rouhani, M.~Oudich, S.~E. Jallal, J.~C. Beugnot, J.~M. Escalante, A.~Martínez,
\newblock \emph{Nanophotonics} \textbf{2014}, \emph{3}, 6 413 .

\bibitem{BAL14-OPT}
K.~C. Balram, M.~Davan\c{c}o, J.~Y. Lim, J.~D. Song, K.~Srinivasan,
\newblock \emph{Optica} \textbf{2014}, \emph{1}, 6 414.

\bibitem{schneider_strong_2016}
K.~Schneider, P.~Seidler,
\newblock \emph{Optics Express} \textbf{2016}, \emph{24}, 13 13850.

\bibitem{seidler_optimized_2017}
P.~Seidler,
\newblock \emph{Journal of Vacuum Science \& Technology B} \textbf{2017}, \emph{35}, 3 031209.

\bibitem{lee_plasma_1997}
J.~W. Lee, J.~Hong, E.~S. Lambers, C.~R. Abernathy, S.~J. Pearton, W.~S. Hobson, F.~Ren,
\newblock \emph{Plasma Chemistry and Plasma Processing} \textbf{1997}, \emph{17}, 2 155.

\bibitem{shul_high-density_1997}
R.~J. Shul, G.~B. McClellan, R.~D. Briggs, D.~J. Rieger, S.~J. Pearton, C.~R. Abernathy, J.~W. Lee, C.~Constantine, C.~Barratt,
\newblock \emph{Journal of Vacuum Science \& Technology A} \textbf{1997}, \emph{15}, 3 633.

\bibitem{smolinsky_plasma_1981}
G.~Smolinsky, R.~P. Chang, T.~M. Mayer,
\newblock \emph{Journal of Vacuum Science and Technology} \textbf{1981}, \emph{18}, 1 12.

\bibitem{pearton_comparison_1996}
S.~J. Pearton, J.~W. Lee, E.~S. Lambers, C.~R. Abernathy, F.~Ren, W.~S. Hobson, R.~J. Shul,
\newblock \emph{Journal of The Electrochemical Society} \textbf{1996}, \emph{143}, 2 752.

\bibitem{pearton_high_1996}
S.~J. Pearton, J.~W. Lee, E.~S. Lambers, J.~R. Mileham, C.~R. Abernathy, W.~S. Hobson, F.~Ren, R.~J. Shul,
\newblock \emph{Journal of Vacuum Science \& Technology B: Microelectronics and Nanometer Structures Processing, Measurement, and Phenomena} \textbf{1996}, \emph{14}, 1 118.

\bibitem{shul_temperature_1996}
R.~J. Shul, A.~J. Howard, C.~B. Vartuli, P.~A. Barnes, W.~Seng,
\newblock \emph{Journal of Vacuum Science \& Technology A} \textbf{1996}, \emph{14}, 3 1102.

\bibitem{yang_experimental_2007}
S.~H. Yang, P.~R. Bandaru,
\newblock \emph{Materials Science and Engineering: B} \textbf{2007}, \emph{143}, 1 27.

\bibitem{abdollahi-alibeik_analytical_1999}
S.~Abdollahi-Alibeik, J.~P. McVittie, K.~C. Saraswat, V.~Sukharev, P.~Schoenborn,
\newblock \emph{Journal of Vacuum Science \& Technology A} \textbf{1999}, \emph{17}, 5 2485.

\bibitem{schneider_nih_2012}
C.~A. Schneider, W.~S. Rasband, K.~W. Eliceiri,
\newblock \emph{Nature Methods} \textbf{2012}, \emph{9}, 7 671.

\end{thebibliography}

\end{document}